\documentstyle[sprocl,epsfig]{article}

\newcommand{\pp}{{_{I\hspace{-0.2em}P}}}

\def\be{\begin{equation}}
\def\ee{\end{equation}}
\def\bea{\begin{eqnarray}}
\def\eea{\end{eqnarray}}

\begin{document}
\title{TARGET FRAGMENTATION AND FRACTURE FUNCTIONS.}

\author{ D. DE FLORIAN}

\address{ Theoretical Physics Division, CERN, CH 1211 Geneva 23, Switzerland}

\author{  R. SASSOT }

\address{ Departamento de F\'{\i}sica, 
Universidad de Buenos Aires \\ 
Ciudad Universitaria, Pab.1 
(1428) Bs.As., Argentina}


\maketitle\abstracts{
We analyse recent data on the production of forward neutrons
in deep inelastic scattering at HERA in the framework of a perturbative QCD description for semi-inclusive processes, which includes fracture functions.
 }


In the most naive quark-parton model picture, the semi-inclusive cross section for the production of a hadron $h$ from  the deep inelastic scattering of charged leptons is usually taken to be given by \cite{aemp}:
\begin{eqnarray}
 \frac{d\sigma^h_p}{dx\,dy\,dz_h}  =  \frac{(1+(1-y)^2)}{2y^2}
  \sum_{i=q,\bar q} c_i\,  f_{i/p}(x)\, D_{h/i}(z_h) \, ,
\end{eqnarray}
where, $ f_{i/p}$ is the parton distribution of flavour $i$,  
$D_{h/i}$ is the fragmentation function of a hadron $h$ from a parton $i$, and
$c_{i}=4\pi e_{q_{i}}^2 \alpha^2/x(P+l)^2$. The kinematical variables used to characterize these processes are:
\begin{equation}
x=\frac{Q^2}{2 P\cdot q} , \ \ \ y=\frac{P\cdot q}{P\cdot l} ,\ \ \ 
z_h=  \frac{P \cdot h}{p \cdot q} =  \frac{E_h }{E_p (1-x)} \frac{1-\cos \theta_h}{2} \, ,
\end{equation}
where $q$ is the transferred momentum $(-q^2=Q^2)$, $l$ and $P$ are the incoming lepton
and proton momenta respectively. $E_h$, $E_p$ and $ \theta_h$ are the produced hadron and target nucleon energies, and the angle between the hadron and the target in the centre of mass of the virtual photon-proton system, respectively.

Although next to leading order corrections to this cross section are also well known, and have been shown to give a very good description of the so-called current fragmentation region ($\theta_h>\pi/2$), the target fragmentation region, which corresponds to $\theta_h=0$  ($z_h=0$), cannot be described with this simplified picture.  First of all, it is easy to see that, at the lowest order, hadrons can only be produced antiparallel to the target nucleon  ($\theta_h=\pi$), excluding the forward configurations. On the other hand, going  to next to leading order, the corrections to the cross section develop divergences  proportional to $1/z_h$, related to soft emission ($E_h=0$), and also to collinear configurations where hadrons are produced in the direction of the remnant target ($\theta_h=0$).  Since at lowest  order hadrons cannot be produced in that direction, it is not possible to factorize the divergence, as usual, into parton distributions and fragmentation functions. 
Then, in order to  describe hadrons produced in the target fragmentation region even at the lowest order, and also to be able to perform at higher orders a consistent factorization   of divergences originated in the current fragmentation region,  a new distribution has to be introduced, the so-called fracture functions, $M_{i,h/N}(x,  (1-x) z )$ \cite{trenvene,graudenz}. These distributions represent  the probability of finding a parton of flavour $i$ and a hadron $h$ in the target $N$ (here  $z = E_h/E_p (1-x)$). 

Therefore the complete leading order expression for the cross section becomes
\begin{eqnarray}
 \frac{d\sigma^h_p}{dx\,dy\,dz} & = & \frac{(1+(1-y)^2)}{2y^2}
  \sum_{i=q,\bar q}  c_i\,  \left[ f_{i/p}(x)\, D_{h/i}(z) \right. \nonumber \\
& +& \left.   (1-x)\, M_{i,h/p} (x, (1-x) z) \right] \, .
\end{eqnarray}
Higher order corrections to this kind of cross section  can be found in refs. \cite{graudenz,npb1,npb2}.

The scale dependence of fracture functions at $\cal{O}$($\alpha_s$) is driven by two kinds of processes, which contribute to the production of hadrons in the remnant target direction: the emission of collinear partons from those found in the target (the usual source of scale dependence of parton distributions, often called homogeneous evolution), and those where partons  radiated from the one to be struck by the virtual probe, fragment into the measured hadron (the so-called inhomogeneous term). These two contributions lead at leading order to the following equation:
\begin{eqnarray}
\frac{\partial}{\partial \log Q^2} M_{i,h/p}\left( \xi,\zeta, Q^2\right) = \frac{\alpha_s(Q^2)}{2\pi} \int_{\xi/(1-\zeta)}^1 \frac{du}{u} \, P^i_j (u)  \, M_{j,h/N}\left( \frac{\xi}{u},\zeta, Q^2\right) \nonumber \\
+ \frac{\alpha_s(Q^2)}{2\pi}\int^{\xi/(\xi+\zeta)}_\xi \frac{du}{\xi(1-u)} \,\hat{P}^{i,l}_j (u) \, f_{j/p} \left(\frac{\xi}{u},Q^2\right)\, D_{h/l} \left(\frac{\zeta u}{\xi (1-u)},Q^2\right) ,
\end{eqnarray}    
 where  $P^i_j (u)$ and $\hat{P}^{i,l}_j (u)$ are the regularized  and real   Altarelli-Parisi splitting functions, respectively.

Recently, the ZEUS Collaboration has measured DIS events identifying high-energy neutrons in the final state \cite{zeusn},  at very small angles  with respect to the proton direction ($\theta_{lab} \leq 0.75 $ mrad), in the kinematical range given by $3\times 10^{-4}< x < 6\times 10 ^{-3}$, $10< Q^2<100$ GeV$^2$ and high $x_L  \simeq z(1-x)  > 0.30$. The ZEUS Collaboration have reported that events with $x_L\geq 0.50$ represent a substantial fraction (of the order of  10\%) of DIS events. In the framework of a picture for semi-inclusive processes including fracture functions, as the one outlined above, the ZEUS findings can be represented
(at LO) by
 \begin{eqnarray}
 \frac{\int_{0.50}^{1-x} \frac{d\sigma^h_p}{dx\,dy\,dx_L}dx_L }  {\frac{d\sigma_p}{dx\, dy} }\equiv \frac{\int_{0.50}^{1-x}
M_2^{n/p}\left(x,\,x_L,\,Q^2  \right)dx_L }{F_2^p \left(x,Q^2\right)} \, .
\end{eqnarray}
   In eq.(5) we have also defined the equivalent to $F_2$ for fracture functions:
$  M_2^{n/p}( x,\, x_L,\, Q^2 ) \equiv  x \sum_i e_i^2 \,  M_{i,n/p} (x, x_L, Q^2)$  
 and we have made explicit the integration over a finite (measured) range of $x_L$. 

In fig. 1 we show the experimental outcome for this fracture function (as defined in eq. (5) and  at $Q^2=10$ GeV$^2$, taking advantage of the negligible $Q^2$ dependence of the data), and we compare it to $F_2^p$ and $F_L^p$.
We also show the contribution to the same observable coming from  current fragmentation processes,  which is pure NLO and  it is about 8 orders of magnitude smaller than the experimental data.  

Fracture functions, as  parton distributions in general, are essentially of a non-perturbative nature and have to be extracted from experiment. However, their close relation with fragmentation and structure functions allows in certain extreme cases a model estimate for them, which can then be compared with actual measurements and evolved with the corresponding evolution equations. As an example, recently, a very sensible model estimate for the production of forward hadrons
in DIS  \cite{niko}, exploiting the idea of non-perturbative 
Fock components of the nucleon has been proposed. In this approach the semi-inclusive DIS cross sections, and through them the corresponding fracture functions at a certain
input scale $Q_{0}^2$,  can be interpreted as the product of a flux of neutrons in the proton (integrated over $p_T^2$) times the structure function of the pion exchanged between them, i.e.
\begin{eqnarray}
M_2^{n/p}\left( x,\, x_L ,Q^2_0 \right) \simeq \phi_{n/p} (x_L) \, F_2^{\pi^+} \left(\frac{x}{1-x_L},Q^2_0\right) \, .
\end{eqnarray} 
Using a non-perturbative computation of the flux, which is in very good agreement with experimental data on high energy neutron and $\Delta^{++}$ production in hadron-hadron collisions  and a parametrization for the pion structure function \cite{grvpion}, in fig. 1 we  make a comparison between the model estimate and the data for $Q^2_0$ = 10 GeV$^2$, finding a remarkable agreement between them.

\setlength{\unitlength}{1.mm}
\begin{figure}[htb]
\begin{picture}(60,60)(0,0)
\put(30,-3){\mbox{\epsfxsize6.8cm\epsffile{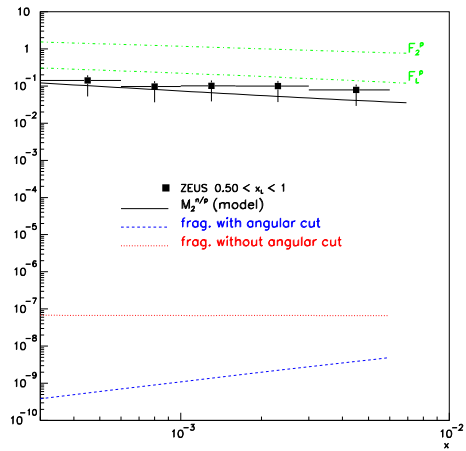}}}
\end{picture}
\caption{ Fracture function of neutrons in protons as measured by ZEUS compared to  the model prediction and current fragmentation contributions \label{fig:radish}}
\end{figure}

The   success of the model estimate encourages us to go further and use the functional dependence of fracture functions, induced by the model and corroborated by the data,  to analyse also the $Q^2$ dependence.
 
We first analyse the more familar process of neutron production. Since the probability of current parton fragmentation into a neutron (given by fragmentation functions) is comparatively small with respect to that of processes originated in the target (fracture functions), no significant effects are expected in the scale evolution arising from the inhomogeneous term in this process. The evolution is mainly driven by the usual homogeneous term of the 
evolution equations leading to an almost constant ratio between the number of neutron tagged events and that of all DIS events, as observed by ZEUS.

However, the scale dependence induced in the cross section for the production of pions, at least in the kinematical region of very small $x$ and small $x_L$,
can be considerably affected by the inhomogeneity, given that soft pions are produced more copiously  from quarks than  from neutrons. In order to analyse these features of the evolution, we estimate the proton to pion fracture function at some input scale $Q^2_0$  using the same ideas formerly applied to neutron production, and noticing that the flux  can be straightforwardly obtained from the one used in the last section by means of  the crossing relation $\phi_{\pi^+/p} (x_L) = \phi^{\pi^+}_{n/p} (1 - x_L)$.
Then, the proton to pion fracture function can be approximated by
\begin{equation}
M_2^{\pi^+/p}\left( x,\, x_L , \,Q^2_0  \right) \simeq \phi_{\pi^+/p} (x_L) \, F_2^{n} \left(\frac{x}{1-x_L}, Q^2_0\right) \, .
\end{equation} 
In fig. 2a we show the model estimates for proton to pion fracture functions (taking   $Q^2_0= 4$ GeV$^2$), integrated over two different bins of $x_L$, compared with the contribution coming from the current fragmentation processes. We assume here the same restrictions as in the data from the ZEUS Collaboration for neutron production. 

\begin{figure}[htb]
\begin{picture}(80,80)(0,0)
\put(10,37){\mbox{\epsfxsize4.9cm\epsffile{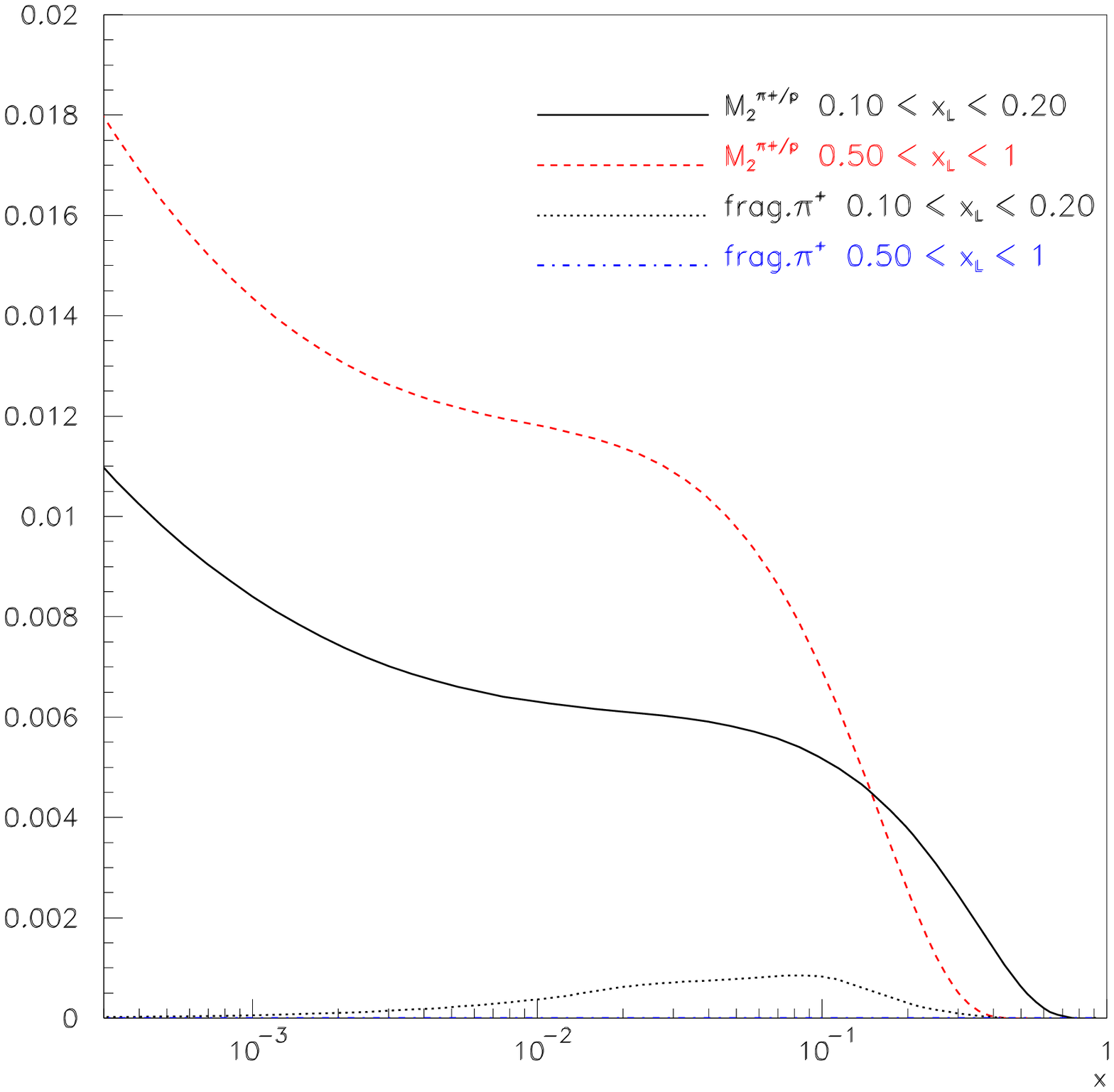}}}
\put(60,37){\mbox{\epsfxsize4.9cm\epsffile{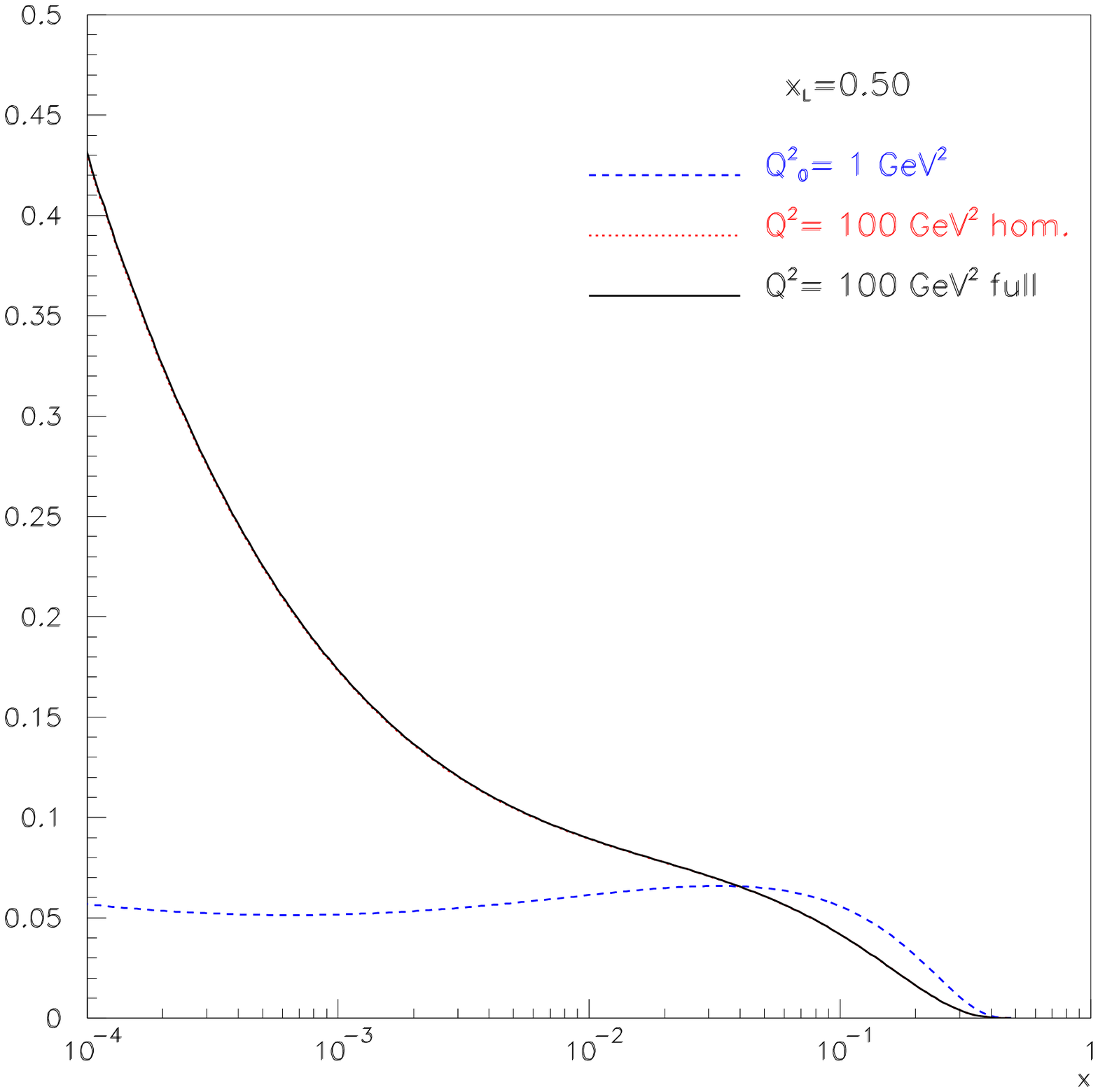}}}
\put(10,-6){\mbox{\epsfxsize4.9cm\epsffile{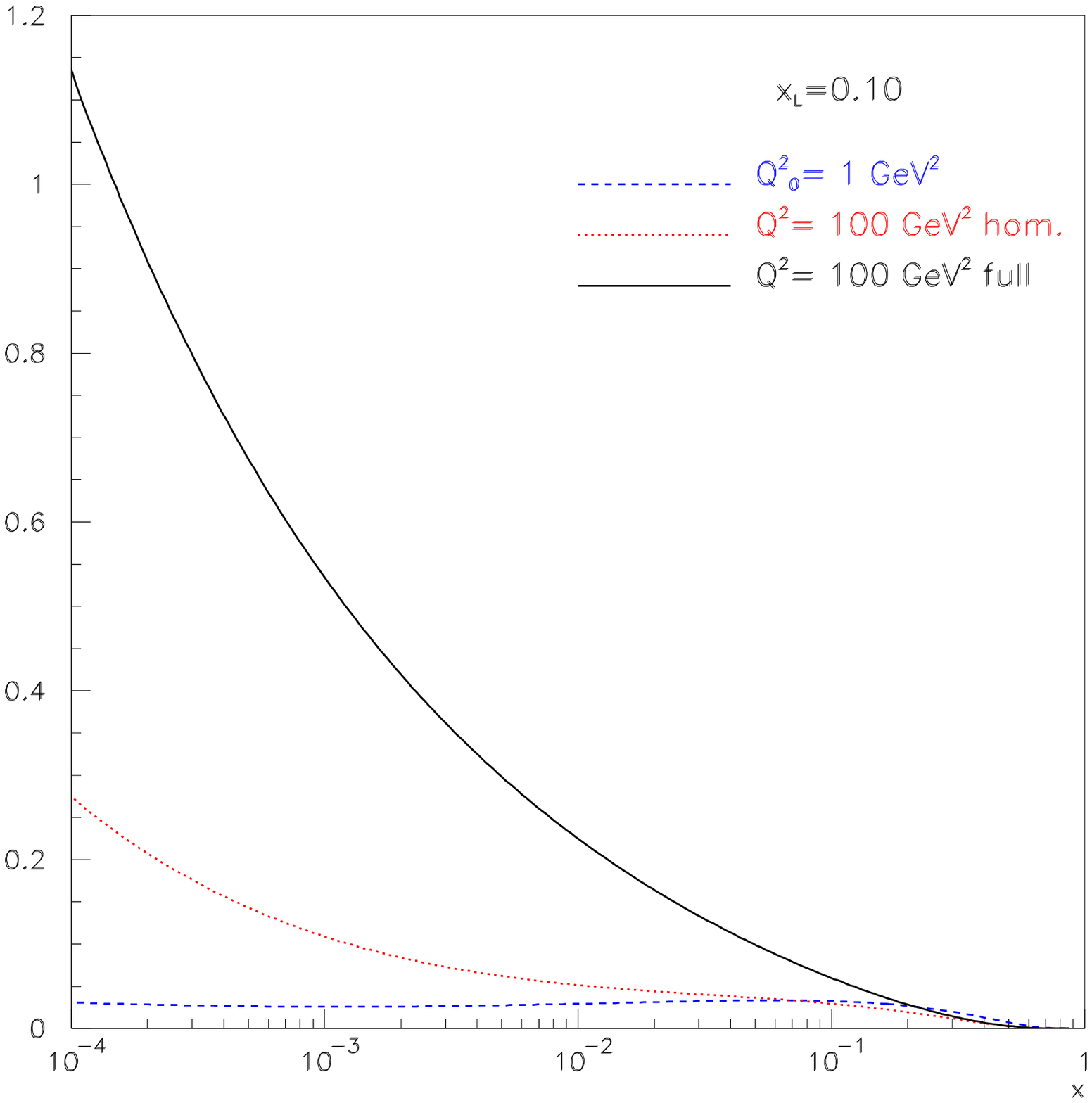}}}
\put(60,-6){\mbox{\epsfxsize4.9cm\epsffile{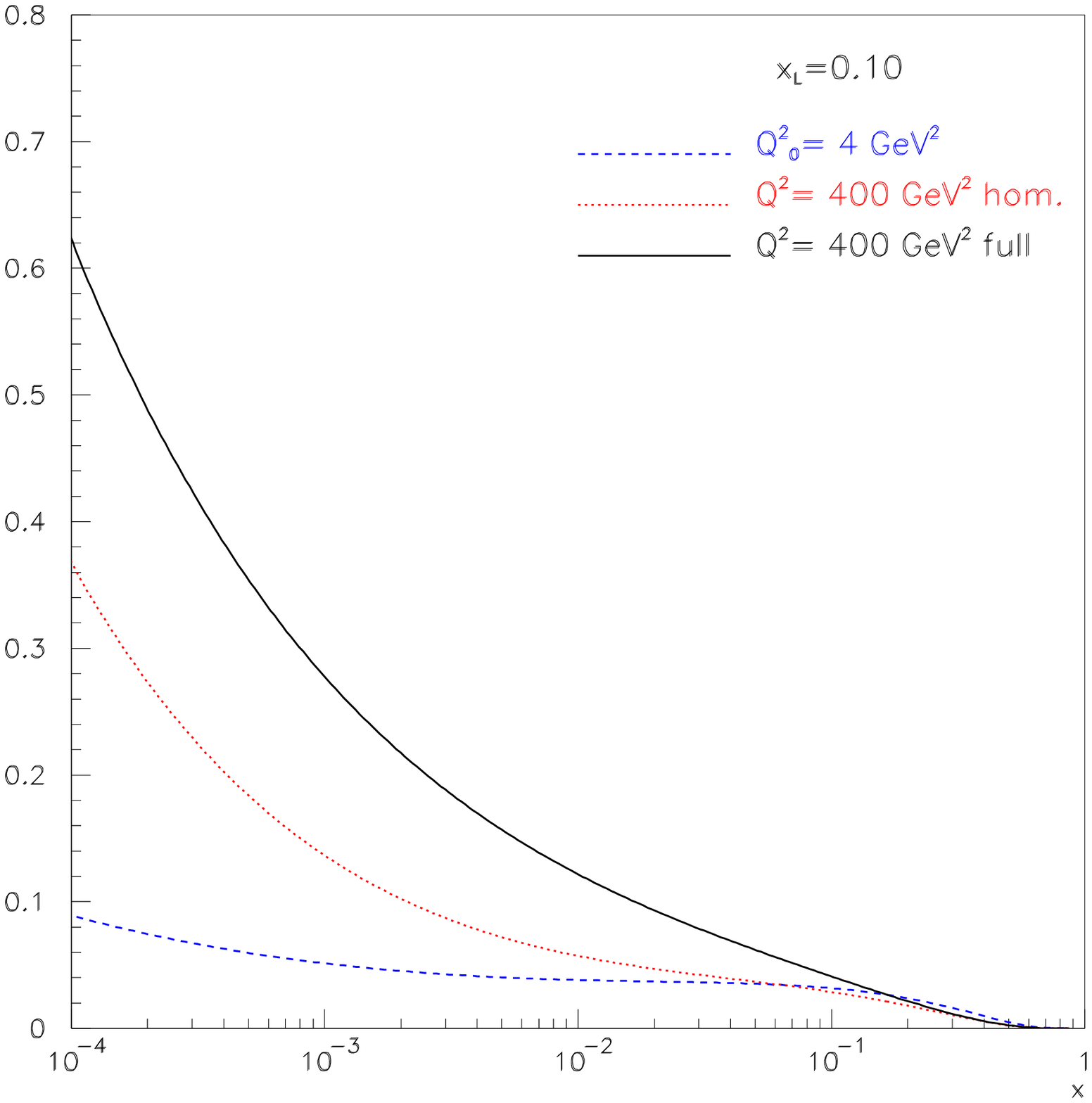}}}
\end{picture}
\caption{a) Prediction for the fracture function of $\pi^+$ in protons for two different bins of $x_L$ and the current fragmentation contribution. 
b) Evolution of the fracture function of $\pi^+$ in protons for $x_L=0.50$ and $Q^2_0=1$ GeV$^2$, c)
$x_L=0.10$, and
d) $Q^2_0=4$ GeV$^2$
\label{fig:radish}}
\end{figure}

Of course, the model is not expected to work over the whole kinematical region and, in fact,  any deviation from the scale dependence implied in eq. (8) (note that the flux is assumed to be $Q^2$ independent) would show the breakdown of the approximated factorization hypotesis. However, the ansatz in eq. (8) can be taken as an effective relation, valid at some initial value of $Q^2_0$, for which the estimated flux is adequate, and therefore provides a sensible input distribution.
As usual, the correct scale dependence is that given by the evolution equations for fracture functions, and that is the aim of our next step. 

In order to study the effect of the inhomogeneity in the evolution we take different values of $x_L$, and keep them fixed while we analyse the $x$ and $Q^2$ dependence of fracture functions induced by  both the homogeneous and the
complete evolution equations.
In fig. 2b we show the result of an evolution from $Q^2_0=1$ GeV$^2$  at $x_L=0.50$. Both solutions, the homogeneous (dotted line) and the complete (solid line) are superimposed,   the difference being less than $0.1\%$. 
 These behaviours are perfectly compatible with the results obtained by the ZEUS Collaboration in the case of neutron production (where the inhomogeneity   contributes about 10 times less) in the same kinematical region, where no difference has been found in the evolution between $F_2$ and $M_2$.

However, for smaller $x_L$ the situation is completely different. As the fragmentation function increases with lower values of the argument, the inhomogeneous contribution becomes much more relevant and its effect in the evolution is sizeable.
In fact,  fig. 2c shows the evolution result for $x_L=0.10$ and $Q^2_0=1$ GeV$^2$, where the full evolution results outsize the homogeneous one by a factor of 4 at small $x$. These corrections are smaller if the ansatz of eq. (8) is assumed to be valid at values of $Q^2_0=4$ GeV$^2$ (fig. 2d)  but  still remain considerable.

Concluding, in this paper we have analysed recent experimental data on the production of
forward neutrons in DIS in terms of fracture functions, finding that the main
features of the data can be fairly reproduced by this perturbative QCD
approach, once a  non-perturbative model estimate for the input fracture
functions is given. Studying the evolution properties of these fracture functions in the specific case of forward pions in the final state, we have
found that the effects of the inhomogeneous term in the evolution equations are large and measurable, particularly in the kinematical region of very small $x$    and small $x_L$.
These effects are negligible for large values of $x_L$, justifying the use of the usual homogeneous Altarelli-Parisi equations for, as an example, the $t$-integrated diffractive structure function $F_2^{D(3)}( x_\pp , \beta, Q^2)$, which is just the fracture function of protons in protons   $M_2^{p/p}\left( \beta x_\pp ,\,( 1- x_\pp), Q^2  \right)$ \cite{grauven}.

\end{document}